\documentclass[twocolumn,showpacs,aps,prl,superscriptaddress]{revtex4}

\usepackage{graphicx}
\usepackage{dcolumn}
\usepackage{amsmath}
\usepackage{epsfig}
\usepackage{bm}

\def\kpmumu{K^+ \rightarrow \pi^+ \mu^+ \mu^-}
\def\kmmumu{K^- \rightarrow \pi^- \mu^+ \mu^-}

\def\kp3pi{K^+ \rightarrow \pi^+ \pi^+ \pi^-}
\def\km3pi{K^- \rightarrow \pi^- \pi^- \pi^+}
\def\kpm3pi{K^{\pm} \rightarrow \pi^{\pm} \pi^+ \pi^-}
\def\kpi{K^{\pm}_{\pi 3}}
\def\kmu{K^{\pm}_{\pi \mu \mu}}
\def\rkppmm{\Gamma(K^+ \rightarrow \pi^+\mu^+\mu^-)}
\def\rkmpmm{\Gamma(K^- \rightarrow \pi^-\mu^+\mu^-)}
\def\rkpmpmm{\Gamma(K^{\pm} \rightarrow \pi^{\pm}\mu^+\mu^-)}

\def\rkpmall{\Gamma(K^{\pm} \rightarrow all)}

\def\Rkppmm{\Gamma(K^+_{\pi\mu\mu})}
\def\Rkmpmm{\Gamma(K^-_{\pi\mu\mu})}

\def\Delkpm{\Delta(K^{\pm}_{\pi\mu\mu})}
\def\kpmu{K^+_{\pi\mu\mu}}
\def\kmmu{K^-_{\pi\mu\mu}}
\def\kpmmu{K^{\pm}_{\pi\mu\mu}}

\def\kpmpi{K^{\pm}_{{\pi}3}}
\def\nokpmu{N^{obs}_{K^+_{\pi\mu\mu}}}
\def\nokmmu{N^{obs}_{K^-_{\pi\mu\mu}}}

\def\nokppi{N^{obs}_{K^+_{\pi{3}}}}
\def\nokmpi{N^{obs}_{K^-_{\pi{3}}}}

\begin{document}

\title{Observation of the Decay $K^{-} \rightarrow \pi^{-} \mu^{+} \mu^{-}$
and Measurements of the Branching Ratios for
$K^{\pm} \rightarrow \pi^{\pm} \mu^{+}\mu^{-}$}

\affiliation{Institute of Physics, Academia Sinica, Taipei 11529, 
             Taiwan, Republic of China}
\affiliation{University of California, Berkeley, California 94720}
\affiliation{Fermi National Accelerator Laboratory, Batavia, Illinois 60510} 
\affiliation{University of Guanajuato, 37000 Leon, Mexico}
\affiliation{Illinois Institute of Technology, Chicago, Illinois 60616}
\affiliation{University of Lausanne, CH-1015 Lausanne, Switzerland}
\affiliation{Lawrence Berkeley National Laboratory, Berkeley, California 94720}
\affiliation{University of Michigan, Ann Arbor, Michigan 48109}
\affiliation{University of South Alabama, Mobile, Alabama 36688}
\affiliation{University of Virginia, Charlottesville, Virginia 22904}

\author{H.K.~Park}
\affiliation{University of Michigan, Ann Arbor, Michigan 48109}
\author{R.A.~Burnstein}
\affiliation{Illinois Institute of Technology, Chicago, Illinois 60616} 
\author{A.~Chakravorty}
\affiliation{Illinois Institute of Technology, Chicago, Illinois 60616}   
\author{A.~Chan}
\affiliation{Institute of Physics, Academia Sinica, Taipei 11529, Taiwan, 
             Republic of China}
\author{Y.C.~Chen}
\affiliation{Institute of Physics, Academia Sinica, Taipei 11529, Taiwan, 
Republic of China}
\author{W.S.~Choong}
\affiliation{University of California, Berkeley, California 94720}
\affiliation{Lawrence Berkeley National Laboratory, Berkeley, 
             California 94720}
\author{K.~Clark}
\affiliation{University of South Alabama, Mobile, Alabama 36688} 
\author{E.C.~Dukes}
\affiliation{University of Virginia, Charlottesville, Virginia 22904}
\author{C.~Durandet}
\affiliation{University of Virginia, Charlottesville, Virginia 22904}
\author{J.~Felix}
\affiliation{University of Guanajuato, 37000 Leon, Mexico}
\author{G.~Gidal}
\affiliation{Lawrence Berkeley National Laboratory, Berkeley, 
             California 94720}
\author{P.~Gu}
\affiliation{Lawrence Berkeley National Laboratory, Berkeley, California 94720}
\author{H.R.~Gustafson}
\affiliation{University of Michigan, Ann Arbor, Michigan 48109}
\author{C.~Ho}
\affiliation{Institute of Physics, Academia Sinica, Taipei 11529, Taiwan, 
             Republic of China}
\author{T.~Holmstrom}
\affiliation{University of Virginia, Charlottesville, Virginia 22904}
\author{M.~Huang}
\affiliation{University of Virginia, Charlottesville, Virginia 22904}
\author{C.~James}
\affiliation{Fermi National Accelerator Laboratory, Batavia, Illinois 60510}
\author{C.M.~Jenkins}
\affiliation{University of South Alabama, Mobile, Alabama 36688}
\author{D.M.~Kaplan}
\affiliation{Illinois Institute of Technology, Chicago, Illinois 60616}
\author{L.M.~Lederman}
\affiliation{Illinois Institute of Technology, Chicago, Illinois 60616}
\author{N.~Leros}
\affiliation{University of Lausanne, CH-1015 Lausanne, Switzerland}
\author{M.J.~Longo}
\affiliation{University of Michigan, Ann Arbor, Michigan 48109}
\author{F.~Lopez}
\affiliation{University of Michigan, Ann Arbor, Michigan 48109}
\author{L.~Lu}
\affiliation{University of Virginia, Charlottesville, Virginia 22904}
\author{W.~Luebke}
\affiliation{Illinois Institute of Technology, Chicago, Illinois 60616}
\author{K.B.~Luk}
\affiliation{University of California, Berkeley, California 94720}
\author{K.S.~Nelson}
\affiliation{University of Virginia, Charlottesville, Virginia 22904}
\author{J.P.~Perroud}
\affiliation{University of Lausanne, CH-1015 Lausanne, Switzerland}
\author{D.~Rajaram}
\affiliation{Illinois Institute of Technology, Chicago, Illinois 60616}
\author{H.A.~Rubin}
\affiliation{Illinois Institute of Technology, Chicago, Illinois 60616}
\author{P.K.~Teng}
\affiliation{Institute of Physics, Academia Sinica, Taipei 11529, Taiwan, 
             Republic of China}
\author{J.~Volk}
\affiliation{Fermi National Accelerator Laboratory, Batavia, Illinois 60510}
\author{C.~White}
\affiliation{Illinois Institute of Technology, Chicago, Illinois 60616}
\author{S.~White}
\affiliation{Illinois Institute of Technology, Chicago, Illinois 60616}
\author{P.~Zyla}
\affiliation{Lawrence Berkeley National Laboratory, Berkeley, California 94720}

\collaboration{HyperCP Collaboration}
\noaffiliation

\date{\today}

\begin{abstract}
\vspace{0.1in}
Using data collected with the HyperCP (E871) spectrometer 
during the 1997 fixed-target run at Fermilab, we report the first observation
of the decay $\kmmumu$ and new measurements of 
the branching ratios for $K^{\pm} \rightarrow \pi^{\pm} \mu^{+}\mu^{-}$. 
By combining the branching ratios for the decays $\kpmumu$ and $\kmmumu$, 
we measure 
$\rkpmpmm{/}\rkpmall = 
(9.8\pm1.0\pm0.5){\times}10^{-8}$.
The CP asymmetry between the rates of the two decay modes is 
$[\rkppmm - \rkmpmm]/[\rkppmm + \rkmpmm] = 
-0.02\pm0.11\pm0.04$.
\end{abstract}


\maketitle

The rare decay of charged $K$ mesons to a pion and
a lepton pair ($K^{\pm}_{\pi l l}$, where $l$ = $e$ or $\mu$)
can be used to study flavor-changing neutral currents as a 
higher-order process in the standard model and to explore new physics.
However, to realize these goals, it is necessary to understand the 
dominant $K^{\pm} \rightarrow \pi^{\pm} \gamma^{*}$ radiative transition which 
involves long-distance hadronic effects.
Since it is difficult to calculate such effects, 
chiral perturbation theory (ChPT)  
including electroweak interactions has been applied to
$K^{+}_{\pi l l}$ \cite{ecker}.   
In a recent model-independent analysis with ChPT at  
$\mathcal{O}$$(p^{6})$ \cite{modin}, 
the $K^{+}_{\pi l l}$ decay rate and the form factor characterizing 
the dilepton invariant-mass spectrum are calculated 
in terms of two parameters $a_{+}$ and $b_{+}$. 
Determining these parameters from the measured branching ratio and the 
dilepton mass spectrum of the $K^{+}_{\pi e e}$ decay, 
this analysis predicts the ratio 
$\mathcal{R}$ = $B(K^{+}_{\pi\mu\mu})/B(K^{+}_{\pi e e})$ 
to be greater than 0.23 and an increase in 
the CP asymmetry between the decay rates of 
$K^{+}_{\pi l l}$ and $K^{-}_{\pi l l}$ compared to the leading-order 
estimation in the chiral expansion \cite{cpest}. 
The resulting expected CP asymmetry is $\sim{10^{-5}}$ \cite{modin,cpest}.

The $K^{+}_{\pi e e}$ decay  has been studied by 
several experiments \cite{kee1,kee2,kee3}, and 
the Particle Data Group (PDG) has compiled a mean branching ratio  
$B(K^{+}_{\pi e e})=(2.88 \pm 0.13) \times 10^{-7}$ \cite{pdg1}. 
The $K^{+}_{\pi \mu \mu}$ decay was first observed by 
the E787 Collaboration at the Brookhaven National Laboratory (BNL).
With 13 fully reconstructed 3-track and 196 partially reconstructed 2-track
events, they determined  
$B(K^{+}_{\pi\mu\mu})$ = 
[5.0 $\pm$ 0.4(stat) $\pm$ 0.7(syst) $\pm$ 0.6(theor)] 
$\times 10^{-8}$ \cite{kmumu1}. 
The E865 Collaboration at BNL has subsequently 
observed 430 fully reconstructed $K^{+}_{\pi\mu\mu}$ events and measured 
$B(K^{+}_{\pi\mu\mu})$ = 
[9.22 $\pm$ 0.60(stat) $\pm$ 0.49(syst)] $\times 10^{-8}$ 
\cite{kmumu2}. The discrepancy between these two experimental results 
is more than three standard deviations and is not understood. 
In addition the value of $\mathcal{R}$ obtained with the E787 result 
and the average value of $B(K^{+}_{\pi e e})$  
is inconsistent with the prediction and is difficult to accommodate 
within the standard model~\cite{modin}.
It is thus important to resolve the discrepancy in the 
$B(K^{+}_{\pi\mu\mu})$ measurements.

\begin{figure}[h]
\centerline{\psfig{figure=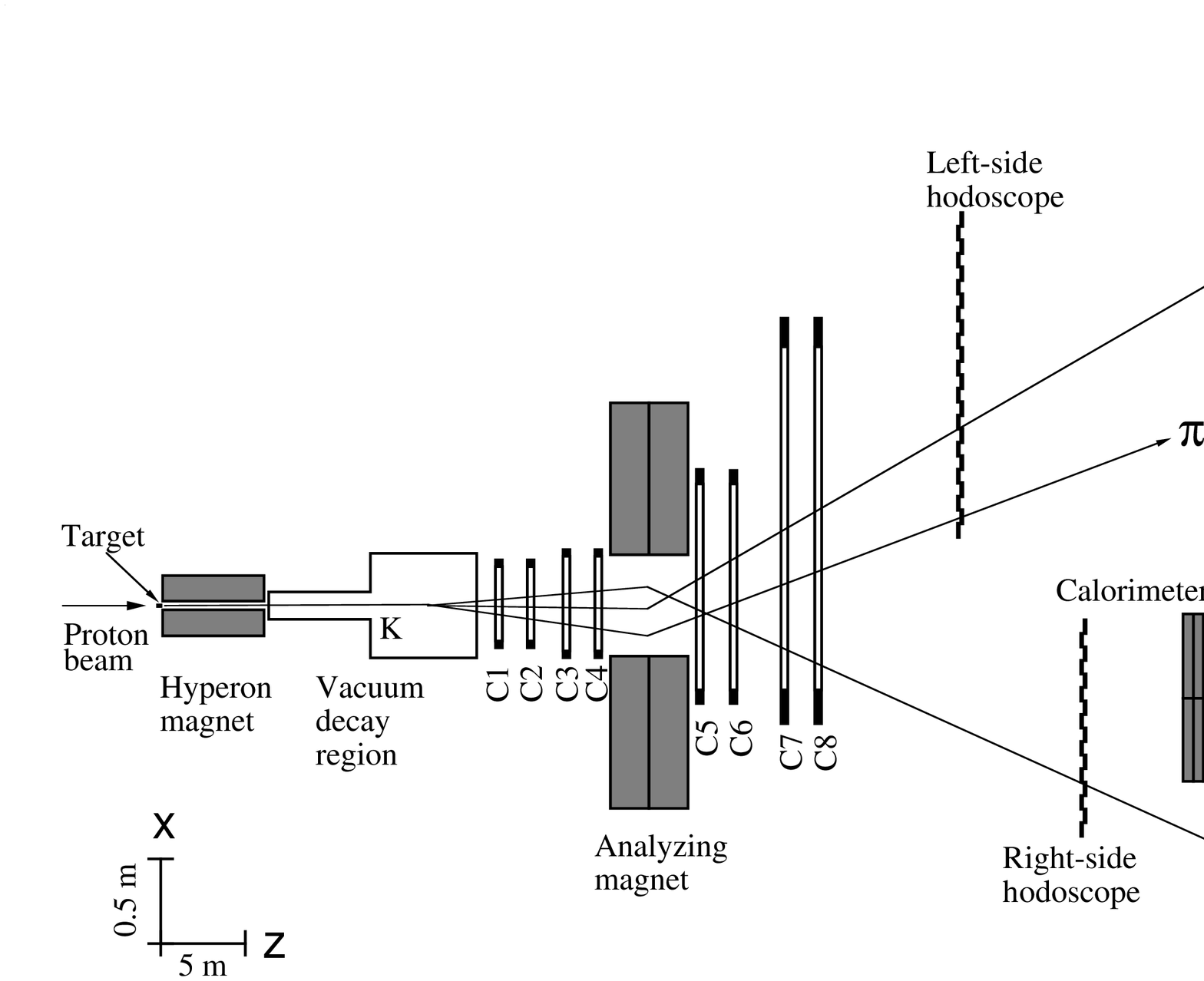,width=87mm}}
\caption{Plan view of the HyperCP spectrometer.
 The length (width) of the spectrometer is about 62 m (4 m).}
\label{fig:spect}
\end{figure}

In this Letter we present results on the first observation of 
$\kmmumu$ decay, the branching ratios of $K^{\pm}_{\pi \mu \mu}$  
decays, and a limit on their CP asymmetry.
The HyperCP (E871) experiment in the Meson Center beam line of Fermilab,
which was designed primarily to study CP violation 
in $\Xi^-/\overline{\Xi}$$^+$ and 
$\Lambda$/$\overline{\Lambda}$ decays, took data in 1997 and 1999.
We report here on an analysis of the 1997 data.  
A plan view of the spectrometer is shown in Fig.~\ref{fig:spect};
a detailed description is given elsewhere \cite{e871}.

$K^\pm$'s were produced by directing an 800-GeV/$c$ proton 
beam, with a typical intensity of $1.2 \times 10^{11}$ 
protons per 20-second spill, onto a 2-mm-square copper target.
A target length of 22 (60)~mm was used when taking 
positive (negative)-secondary-beam data.
The secondary-beam momentum was selected by a curved collimator 
embedded in a 6-m-long dipole magnet (Hyperon magnet) with a horizontal 
magnetic field of 1.67~T.  
The mean momentum of the secondaries was about 170~GeV/$c$, and the charge 
of the secondary beam was selected with the sign of the Hyperon magnet field.
The typical secondary-beam rate was 13~MHz at the exit of the collimator. 
Following a 13-m evacuated decay pipe were eight multiwire proportional 
chambers (C1--C8) with wire spacings increasing from 1 to 2~mm.
A dipole magnet (Analyzing magnet) deflected charged particles 
horizontally with a transverse-momentum kick of 1.43~GeV/$c$.  
Particles with the same charge as the secondary beam were 
deflected to the left and those with opposite charge to the right.
At the rear of the spectrometer were two scintillation hodoscopes and
a calorimeter, used for triggering, and a muon detector system. 
The muon system consisted of two similar detectors on either side of the 
secondary beam line. The muon detectors had three layers of 0.81-m-thick 
steel absorber, each layer followed by a tracking station 
consisting of vertical and horizontal planes
of proportional tubes with a pitch of 25.4 mm.
At the rear of the third layer, behind the proportional tubes,
were two orthogonal scintillator hodoscopes used for the muon triggers.

The $\kpmmu$ rates were measured relative to 
the $\kpm3pi$ ($\kpmpi$) rates (the normalization mode).
The trigger for the $K^{\pm}_{\pi 3}$ events was a heavily-prescaled
coincidence of at least one hit in each of 
the left- and right-side hodoscopes.
The trigger for the $\kpmmu$ events, the unlike-sign dimuon trigger,
required the presence of at least one hit in each hodoscope in the muon 
detectors in coincidence with hits in the left- and right-side hodoscopes. 
The sign of the secondary beam was changed every few hours
by reversing the polarities of both the Hyperon and Analyzing magnets.
The typical run cycle was two runs with positive secondary beam 
followed by one run with negative secondary beam. 

Data for the $\kpmmu$ and $\kpmpi$ triggers were 
processed with the same reconstruction program.
A three-track event topology was required.    
The sample of $\kpmmu$ candidates was selected by 
requiring a muon track in each  muon station
and an unlike-sign dimuon trigger.
The muon track was required to have hits in at least two of three muon 
chambers in both $x$ and $y$ views with corresponding in-time hits in 
the muon hodoscope.  The algorithm for muon selection was checked 
using muon calibration data. 
The momentum of accepted muons was greater than 20~GeV/$c$.

Similar selection criteria were applied to both the signal ($\kpmmu$)
and normalization ($\kpmpi$) event samples.
One right-side track and two left-side tracks were required.
Two pions from the $K^{\pm}_{\pi 3}$ decays were required to project to 
the same fiducial areas in the left- and right-side muon hodoscopes 
as the muons from the $K^{\pm}_{\pi \mu \mu}$ decays.
The total momentum of the reconstructed tracks was required to
lie between 120 and 250~GeV/$c$, consistent with the momentum spread 
of the secondary beam. 
To ensure that the kaons came from the target, the combined three-track 
momentum had to point back to within 5~mm of the center of the target in 
both $x$ and $y$. 
The decay vertex of the three tracks was calculated by the method of 
distance of closest approach using the front track segments, and 
was required to be well within the vacuum decay region.

The $\kpmmu$ and $\kpmpi$ event candidates were required to
have a decay topology consistent with a single vertex.
This was done by first fitting the three tracks to a single
vertex using the wire hits in C1--C4 and computing the $\chi^2$ of the fit.
In addition the average separation between the three pairs of tracks
in the $x-y$ plane at the $z$ position of the vertex 
determined by distance of closest approach was measured.
Clean $\kpmpi$ decays were used to study selection cuts 
based on the vertex-constrained-fit $\chi^2$ and the average separation
distance.  Candidates were accepted if $\chi^2/ndf < 2.5$ and 
the average separation distance was less than 2~mm.
These vertex cuts provided strong rejection of backgrounds,
in particular those from two-vertex hyperon decays.
\begin{figure}[h]
\centerline{\psfig{figure=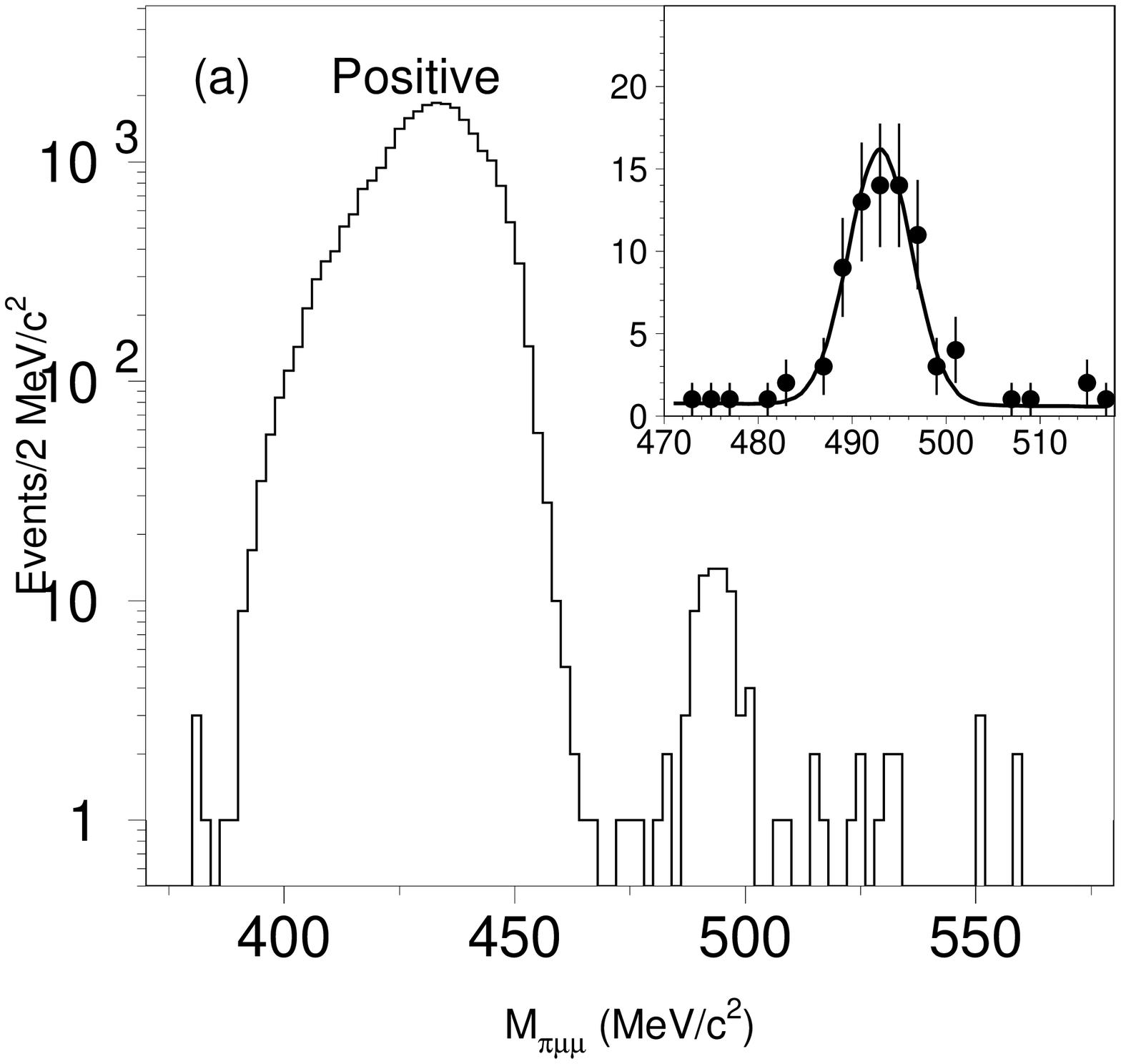,height=60mm,width=80mm}}
\vspace{0.5cm}
\centerline{\psfig{figure=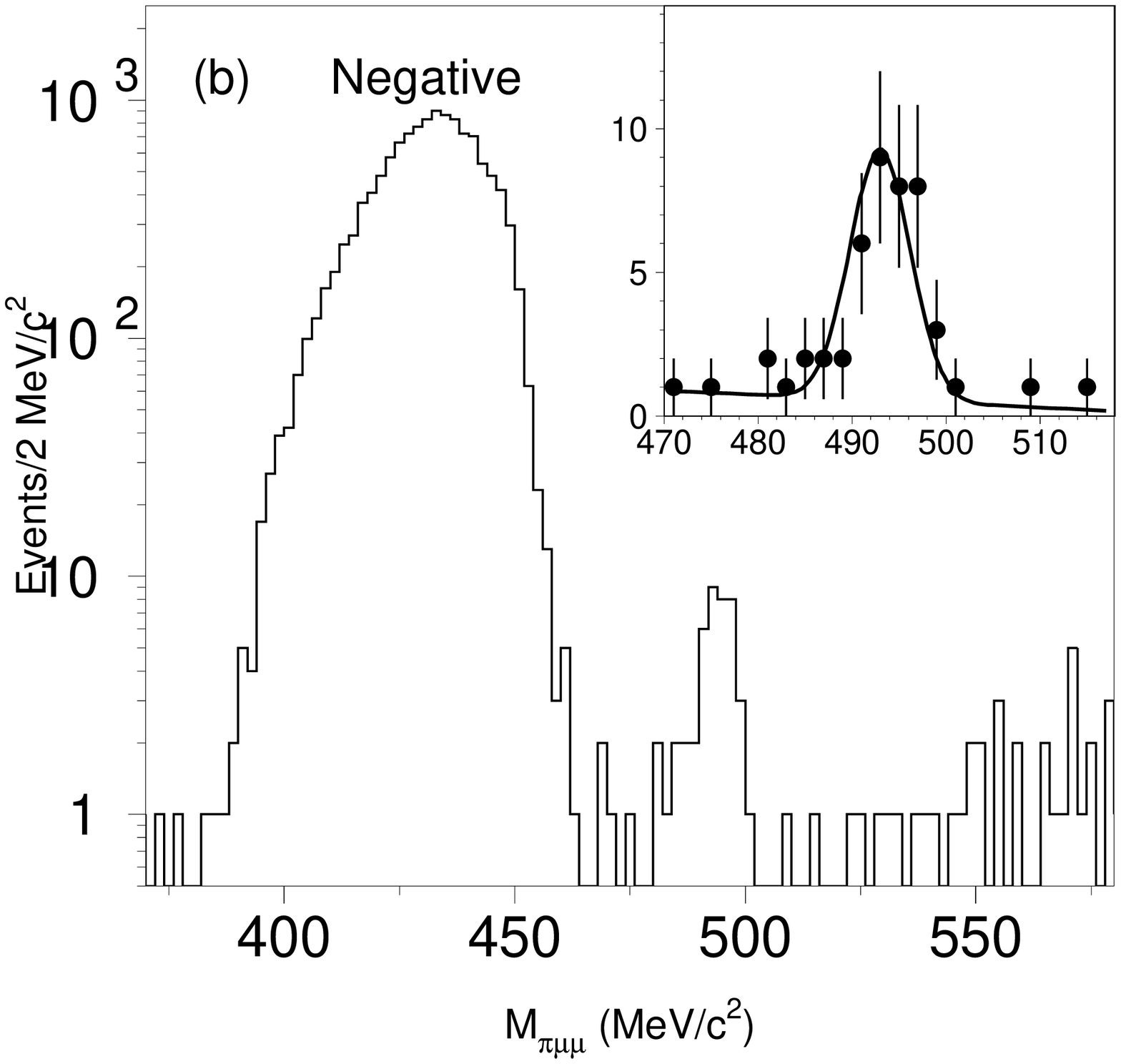,height=60mm,width=80mm}}
\caption{The $\pi\mu\mu$ invariant mass distributions
for (a) positive- and (b) negative-secondary-beam data after cuts.  
The large peak comes from  $K^{\pm}_{\pi 3}$ events with 
in-flight decays of pions and pion punch-through to the muon stations. 
The inset shows the region around the kaon mass 
in more detail; points are data and the solid lines show the results of 
the likelihood fits.}
\label{fig:invmass}
\end{figure}  

For events passing these cuts, the reconstructed three-track invariant 
mass $M_{\pi \mu \mu}$ distributions with a $K^{\pm}_{\pi \mu \mu}$ 
decay hypothesis are shown in Fig.~\ref{fig:invmass}.
The signals in the vicinity of the kaon mass are clear and unambiguous.

To determine the branching ratios $B(\kpmmu) = \rkpmpmm/\rkpmall$ the
following formula was used:
\begin{eqnarray}
B(\kmu)=\frac{N_{\kmu}^{obs}}{200 \cdot N_{\kpi}^{obs}}
        \frac{A_{\kpi}}{A_{\kmu}}
        \frac{\epsilon_{\kpi}}{\epsilon_{\kmu}}
        \frac{ B(\kpi) }{\epsilon_{\mu^+ \mu^-} \cdot \epsilon_{rel}^{trig}},
\label{eq:ratio}
\end{eqnarray}
where the $N^{obs}_{i}$'s are the numbers of observed events,
the $A_{i}$'s are the geometrical acceptances,
the $\epsilon_{i}$'s are the event-selection efficiencies,
$\epsilon^{trig}_{rel}$ is the relative trigger efficiency, 
and 200 is the prescale factor.
The branching ratios $B(K^{\pm}_{\pi3})$ are $(5.59 \pm 0.05)\%$~\cite{pdg1}.
Unbinned maximum-likelihood fits for the events in the
region $470~MeV/c^2 < M_{\pi \mu \mu} < 520~MeV/c^2$ 
were used to estimate the numbers of $\kpmmu$ events, 
where the mass range for the fit
was chosen to minimize the statistical error.  
Gaussian and linear functions were used to fit the signal and background 
respectively.
The numbers of signal events from the fits were 
$\nokpmu = 65.3\pm{8.2}$ and $\nokmmu = 35.2\pm{6.6}$, 
where the errors are statistical.
After background subtraction, the observed numbers 
of normalization events were
$\nokppi = (4.446 \pm 0.010){\times}10^{5}$ and 
$\nokmpi = (2.318 \pm 0.008){\times}10^{5}$, where
the errors are dominated by the background estimation.

Background contributions to the $K^\pm_{\pi\mu\mu}$ samples
from pion punch-through in the muon stations and
in-flight decays of pions in $\kpmpi$ decays and 
from $K^{\pm} \rightarrow \pi^+ \pi^- \mu^{\pm} \nu_{\mu}$ were 
negligible, mainly due to the good resolution of the spectrometer. 
Monte Carlo (MC) studies for these decays showed that the
reconstructed masses with the $\pi \mu \mu$ hypothesis of these modes were well
below 480 MeV/$c^2$.

Many checks to verify the MC simulation were made. 
The kinematic parameters for $K^{\pm}$ production at the target
were tuned separately for the $K^+_{\pi3}$ and $K^-_{\pi3}$
normalization samples.  
The PDG Dalitz parameters for $K^{\pm}_{\pi 3}$ decays \cite{pdg1} 
were used in the MC program to properly simulate this decay. 
The distributions for kaon momenta and vertex positions of 
$\kpmpi$ decays inside the decay pipe, hit positions of pions 
from the decays, and reconstructed masses were compared with data.
They all showed good agreement.

\begin{figure}[h]
\centerline{\psfig{figure=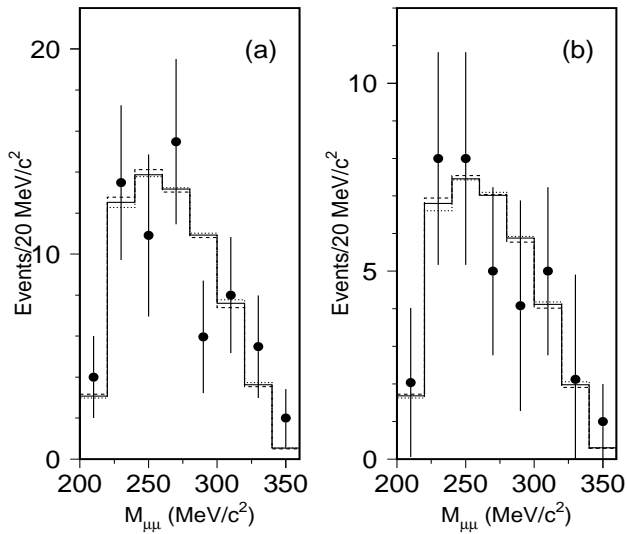,height=70mm,width=82mm}}
\caption{The $\mu^{+} \mu^{-}$ invariant mass after background subtraction
for (a) positive- and (b) negative-secondary-beam data
after cuts. Points are data. The solid, dashed and dotted lines represent 
the simulation result with  $\delta=2.08$, $1.72$ and $2.44$, 
respectively.}
\label{fig:mumass}
\end{figure}

In the simulation of $K^{\pm}_{\pi \mu \mu}$ decays, we assumed 
a vector interaction, based on the results of 
previous experiments \cite{kee3,kmumu2}, and so
the angular distribution of $\mu^+$ in the rest frame of 
the virtual photon was generated according to the expression 
$(m_{\mu}^2 + p_{\mu}^2 \sin^2 \theta)$, where $m_{\mu}$ and $p_{\mu}$
are the mass and momentum of the muon, and
$\theta$ is the angle between the directions of the $\mu^+$ and the pion 
in this frame.   
In addition, we employed a form factor as described below.  
The dilepton invariant mass ($M_{ll}$) spectrum 
from the $K^{\pm}_{\pi l l}$ decays is described 
by the form factor
$\phi(M_{ll}^2)=\phi_{0}(1 + \delta M_{ll}^2/m_{K^+}^2)$
where $\delta$ and $\phi_{0}$ are a slope parameter 
and a constant, respectively, 
and $m_{K^+}$ is the mass of the charged kaon \cite{kee2,kee3}.
To extract $\delta$, we averaged the measured values of the slope 
parameters determined by BNL E777 using $K^{+}_{\pi e e}$ decays~\cite{kee2} 
and by BNL E865 with $K^{+}_{\pi e e}$~\cite{kee3} 
and $K^{+}_{\pi\mu\mu}$ events~\cite{kmumu2}.
We obtained $\delta=2.08 \pm 0.12$, where
the error is only statistical. 
Figure~\ref{fig:mumass} compares data for the dimuon invariant mass for
$K^{\pm}_{\pi \mu \mu}$ decays with simulation results 
for 3 values of $\delta$.

From the MC simulation the geometric acceptance for 
$K^{\pm}_{\pi 3}$ and $K^{\pm}_{\pi \mu \mu}$ decays inside the vacuum
decay region for positive (negative)-secondary-beam data
were estimated to be 94.4\% (94.2\%) and 47.5\% (47.7\%), respectively,
where we have taken into account pion decay in flight 
and pion punch-through to the muon stations. 
Analyses identical to those applied to the data were used for the simulated  
normalization and signal events that passed the trigger requirements.
The event selection efficiencies for  
$K^{\pm}_{\pi 3}$ and $K^{\pm}_{\pi \mu \mu}$ decays were 77.9\% (76.0\%) and
80.3\% (78.3\%) for positive (negative) secondary beam data respectively. 
The difference in efficiencies for $K^{+}$ and $K^{-}$ events 
is primarily due to the slightly different momentum spectra 
for $K^{+}$ and $K^{-}$ production.
To estimate the trigger and dimuon selection efficiencies,
the full data sample was used.
Since the trigger for the normalization mode was common to 
that for the signal mode, we could determine the relative 
trigger efficiency for the signal mode with
respect to the normalization mode.
The relative trigger efficiency $\epsilon_{rel}^{trig}$ was found to be 
$(86.6 \pm 2.7)\%$. The dimuon selection efficiency $\epsilon_{\mu^+ \mu^-}$
was $(93.8 \pm 0.3)\%$.

A summary of the systematic uncertainties is given in Table \ref{tab:syst}.
Systematic uncertainties on the measured branching ratios were minimized
by using the same cuts both for the signal and for the
normalization modes, with the exception of the muon-identification cuts
which were applied to the signal events only.
The major systematic uncertainties were the background estimation
for the signal events and the trigger efficiency.
In order to estimate the uncertainties in the numbers of 
signal events, various mass ranges were used for the fits, 
and fits with constant and quadratic functions for the background were used.
The variations of the beam position at the target and
the magnetic field strength in the Hyperon 
and Analyzing magnets during the run were monitored. 
By varying the assumed beam position, magnetic fields, 
$K^{\pm}$ production parameters, and the decay parameters in the MC, 
we determined the sensitivity
to these parameters of the relative acceptances and efficiencies for the
signal and normalizing modes.
Realistic ranges for these parameters were
then used in computing the MC-dependent systematic uncertainties 
in Table \ref{tab:syst}.
Finally, the effect of the uncertainty in the 
slope parameter $\delta$ was estimated by varying
it by three standard deviations (see Fig.~\ref{fig:mumass}), 
and found to be negligible.

Based on Eq. (\ref{eq:ratio}) we determined 
the branching ratios to be 
$B(\kpmu) = (9.7\pm 1.2\pm 0.4) \times 10^{-8}$
and $B(\kmmu)$ = ($10.0\pm 1.9\pm 0.7) \times 10^{-8}$, 
where the first and second errors are statistical and systematic respectively. 
In terms of the partial decay rates $\Gamma(K^{\pm}_{\pi\mu\mu})$, 
the CP asymmetry in these modes is defined as
\begin{eqnarray}
\Delkpm & = & \frac{\Rkppmm - \Rkmpmm}{\Rkppmm + \Rkmpmm}.
\end{eqnarray}   
From our measurements of $B(K^{\pm}_{\pi\mu\mu})$, we obtained 
$\Delkpm = -0.02\pm 0.11{\rm (stat)} \pm 0.04{\rm (syst)}$.

In conclusion, we have observed both $\kpmumu$ and $\kmmumu$ decays
and measured their branching ratios.
This is the first observation of the $\kmmumu$ decay. 
The CP asymmetry parameter, $\Delkpm$, has been extracted from
these measurements.  
Our result is consistent with no CP violation.
Assuming CP symmetry is valid, 
we can combine  $B(\kpmu)$ and $B(\kmmu)$ to give a result of 
[9.8 $\pm$ 1.0(stat) $\pm$ 0.5(syst)]$\times 10^{-8}$,  
which is 3.2 standard deviations higher than 
the BNL E787 measurement \cite{kmumu1},
but consistent with the BNL E865 result \cite{kmumu2} and,
using the PDG value for $B(K^{+}_{\pi e e})$, is also consistent
with the model-independent analysis for the $\mathcal{R}$ value~\cite{modin}.
  
\begin{table}[htpb]
\centering
\begin{tabular}{lcc}
\hline \hline  
Source       &  $\sigma_{B}/B$ ($\%$) 
             &  $\sigma_{B}/B(K^{\pm}_{\pi\mu\mu})$ ($\%$) \\ \hline 
Beam targeting                                  & 1.1 (0.9)     
                                                & 1.0      \\ 
Magnetic field                                  & 0.5 (0.4)    
                                                & 0.5      \\ 
Trigger efficiency                              & 3.1 (3.1)     
                                                & 3.1      \\
Muon identification                             & 0.3 (0.3)   
                                                & 0.3      \\ 
Background Estimation ($K^{\pm}_{\pi 3}$)       & 0.2 (0.3)    
                                                & 0.2      \\
Background Estimation ($K^{\pm}_{\pi \mu \mu}$) & 2.4 (6.4)    
                                                & 3.6      \\
Data and MC disagreement                        & 0.4 (0.4)    
                                                & 0.4     \\ 
Dalitz parameter                                & 0.3 (0.3)    
                                                & 0.3      \\ 
Slope parameter ($\delta$)                      & 0.2 (0.3)    
                                                & 0.2      \\ 
$B(\kp3pi)$                                     & 0.9 (0.9)    
                                                & 0.9      \\ \hline 
Total                                           & 4.2 (7.3)    
                                                & 5.0      \\ \hline \hline
\end{tabular}
\caption{Summary of estimated systematic uncertainties in the branching
ratios of $B(\kpmu)$, and, in parentheses, $B(\kmmu)$.
The last column gives the systematic errors for the combined branching
ratio, $B(\kpmmu)$.}
\label{tab:syst}
\end{table} 

\begin{acknowledgments}
The authors are indebted to the staffs of Fermilab and
the participating institutions for their vital contributions.
This work was supported by the U.S. Department of Energy and
the National Science Council of Taiwan, R.O.C.
\end{acknowledgments}

\end{document}